\documentclass[aps,prx,showpacs,amsmath,amssymb,amsfonts,lengthcheck,twocolumn,footinbib,floatfix,superscriptaddress,longbibliography]{revtex4-2}
\usepackage[utf8]{inputenc}
\usepackage[T1]{fontenc}
\usepackage{amsmath,amssymb}
\usepackage[colorlinks=true,citecolor=blue,linkcolor=blue,urlcolor=blue]{hyperref}

\usepackage{url}
\usepackage{graphicx}
\usepackage{scrextend}
\usepackage{xcolor}
\usepackage{cleveref}
\usepackage{soul}
\usepackage{bm}

\definecolor{orange}{rgb}{1,0.5,0}

\newcommand{\ef}[1]{#1}

\begin{document}

\title{Long-time diffusion and energy transfer in polydisperse mixtures of particles with different temperatures}
\author{Efe Ilker}
\email{ilker@pks.mpg.de}
\affiliation{Laboratoire Physico-Chimie Curie, Institut Curie, PSL Research University, CNRS UMR 168, Paris, France}
\affiliation{Sorbonne Universités, UPMC Univ. Paris 06, Paris, France}
\affiliation{\ef{Max Planck Institute for the Physics of Complex Systems, 01187 Dresden, Germany}}
\author{Michele Castellana}
\affiliation{Laboratoire Physico-Chimie Curie, Institut Curie, PSL Research University, CNRS UMR 168, Paris, France}
\affiliation{Sorbonne Universités, UPMC Univ. Paris 06, Paris, France}\author{Jean-Fran\c{c}ois Joanny}
\affiliation{Laboratoire Physico-Chimie Curie, Institut Curie, PSL Research University, CNRS UMR 168, Paris, France}
\affiliation{Sorbonne Universités, UPMC Univ. Paris 06, Paris, France}\affiliation{Coll\`{e}ge de France, 11 place Marcelin Berthelot, 75005 Paris, France}
\begin{abstract}
Evidence suggests that the transport rate of a passive particle at long timescales is enhanced due to interactions with the surrounding active ones in a size- and composition-dependent manner. Using a system of particles with different temperatures, we probe these effects in dilute solutions and derive long-time friction and self-diffusion coefficients as functions of volume fractions, sizes and  temperatures of particles in $d=2$ and 3 dimensions. Thus, we model excluded-volume interactions for nonequilibrium systems but also extend the scope to short-range soft potentials and compare our results to Brownian-dynamics simulations. Remarkably, we show that both viscosity and energy flux display a nonlinear dependence on size.  The simplicity of our formalism allows to discover various interesting scenarios that can be relevant for biological systems and active colloids.
 \end{abstract}

\maketitle

\section{Introduction}
A particle in a solvent is constantly bombarded by the solvent molecules (receiving random 
kicks at a rate $\sim \tau_{\rm s}^{-1}$) which push the particle, while the particle dissipates 
the excess energy through dynamical friction also exerted by the solvent  at 
longer timescales. This leads to the well-known Brownian motion  
\cite{einstein1905motion}. The transport is measured by the mean-square displacement 
(MSD) of the particle, which is determined by its diffusion constant $D=T/\zeta$ at time 
periods $t\gg \tau_{\rm s}$. Here, $T$ is the temperature of the solvent, i.e., the ``bath" temperature, 
and $\zeta$ is the friction coefficient of the particle, which depends on the viscosity of 
the solvent and the size of the particle. For finite concentrations of interacting particles, at much longer time spans $t\gg \tau_{\rm c}$
the diffusion constant of a single tagged particle differs from the bare diffusion 
constant $D$, because the particle  
experiences numerous collisions with the surrounding particles at a rate $\sim \tau_{\rm c} ^{-1}$. 
In equilibrium fluids, this effect is a consequence of the effective friction due to the 
excluded-volume interactions \cite{leegwater1992dynamical} and can only reduce the value of 
long-time value diffusion constant $D^{\rm s}$ of the tagged particle, i.e., its self-diffusion coefficient.

By contrast, in an active fluid, a passive tracer particle can gain extra energy through 
interactions with the surrounding active particles 
\cite{wu2000particle,leptos2009dynamics,mino2011enhanced}. The
self-diffusion coefficient is then \ef{determined} by the interplay between the effective friction
discussed above and this energy transfer. In biological systems, crowding, energy transmission and composition are key factors on the transport of material 
\cite{minton2001influence,ellis2001macromolecular,delarue2018mtorc1,vibhute2020transcription,movilla2020two}, and out-of-equilibrium aspects enrich the dynamics. Notably in the cytoplasm, not only the long-time friction scales nonlinearly \cite{kalwarczyk2011comparative,etoc2018non}, but also the energy flux varies with the size of the probe particle \cite{parry2014bacterial}. Disentangling these elements is difficult in general polar active 
models. Yet, we can attempt to establish simpler models which keep the relevant features
of these systems. In particular, activity brings extra persistence with velocity $v$ to the translational 
motion of active particles while the direction is randomized at a rotational timescale 
$\tau_{\rm r}$. At intermediate timecales $\tau_{\rm c}> t \gg \tau_{\rm r}$, this motion appears diffusive with 
\ef{$D\sim v^2\tau_r $}, and can be described by an 
effective temperature, higher than the thermal one. \ef{Similar characteristics can be achieved by phoretic motility in artifical self-propelled particles \cite{illien2017fuelled}.}
While $\tau_{\rm r}$ is very short for biomolecules, it gains importance for \ef{self-propelled particles}. At \ef{timescales $t< \tau_c$, the tagged particle diffusion is only perturbed by hydrodynamic interactions  \cite{cheng2002nature,van1992long}.} A recent study explores how the hydrodynamic field of the microswimmer leads to a ballistic motion of tracer particles at short to intermediate timescales \cite{kanazawa2020}. On the other hand, at long timescales $t\gg \tau_{\rm c}$, \ef{the dynamics converges to diffusive motion where} direct collisions dominate and the effect of the hydrodynamic field is decreased in the long-time self-diffusion coefficient  $D^{\rm s}$ \cite{hanna1982self,thorneywork2017self}. This allows for a consistent separation of timescales to implement a multiple-temperature model for studying the long-time transport properties in out-of-equilibrium systems.

Various studies using particles with different temperatures explored rich phenomena beyond 
equilibrium mixtures: Active/passive phase separation of colloids 
\cite{grosberg2015,weber2016}, polymeric systems \cite{smrek2017,smrek2018,chubak2020emergence}, activity-mediated
interactions \cite{tanaka2017}, interaction-dependent temperatures \cite{wang2019}, diffusion of a passive particle weakly coupled to an active field (and vice versa) \cite{demery2011perturbative} among others. Recently, we constructed a theory of phase separation for these systems, and showed that the three-body correlations lead to non-reciprocal interactions upon coarse-graining \cite{ilker2020phase} and hence a shift from an effective equilibrium construction. This reflects similar governing principles and peculiarities observed in polar active models \cite{solon2018,tjhung2018}. Moreover, the concept of a second temperature successfully describes other systems, ranging from biology to plasma physics \cite{pande2000,exartier1999,cugliandolo2011,abdoli2020correlations,netz2020approach,crisanti2012,chertovich2004,pitaevskii2012,ferriere2001interstellar,wolfire2003}.

In this work, we provide a theoretical and numerical analysis of the long-time behavior of transport properties and energy transfer as functions of sizes, temperatures and concentrations in mixtures of particles with different temperatures. We start from the Langevin dynamics with excluded-volume interactions while neglecting hydrodynamic interactions, and derive the long-time transport coefficients expanded up to first order in concentrations. \ef{In Section \ref{sec2}, we describe the model and highlight the central results of this paper which we compare with the Brownian dynamics simulations. In Section \ref{sec3}, we discuss how modeling of excluded-volume interactions in nonequilibrium systems differs from the equilibrium systems. The detailed derivations are included in Supplemental Material \cite{supplement} which we hope to be useful for further studies in nonequilibrium systems as we extend the scope of microrheology approaches developed in equilibrium systems and simplify the framework. }

\section{Model and Results}\label{sec2}
We consider 
the motion of colloidal particles in $d=2,3$ dimensions following overdamped Langevin dynamics 
\begin{eqnarray}\label{eq:1}
\zeta_{m} \dot{\bf r}_{m}=-\partial_{{\bf r}_m} U_{N} + (2 T_{m} \zeta_{m})^{1/2} {\bm \xi}_{m} (t).
\end{eqnarray}
The total potential energy $U_{N}=\frac{1}2\sum_{i,j}u(|{\bf r}^{(i)}-{\bf 
r}^{(j)}|)+\sum_i u_{\rm v}({\bf r}^{(i)})$ includes both the interactions between 
particles and an external potential which confines our system in a volume $V$. 
The 
position of particle $i$ is denoted by ${\bf r}_{i}$, its friction 
coefficient by $
\zeta_{i}$, its temperature by $T_{i}$ and ${\bm \xi}_{i}(t)$ is a $d-$dimensional  
standard Gaussian white noise vector with
zero mean and
unit variance, i.e., $\langle{\bm \xi}_{i}(t)\cdot{\bm 
\xi}_{j}(t')\rangle=d \, \delta(t-t')\delta_{ij}$.
We consider multiple species of particles. 
Each particle of species $\alpha$ is
in contact with a thermostat at 
temperature $T_{\alpha}$ and friction coefficient $\zeta_{\alpha}$. In general, the 
observables of the multi-particle system can be expressed as a series expansion 
in the particle densities. At first order in  concentrations, the correction 
to isolated particle properties results from the sum of the contributions of all 
two-particle clusters.

For two particles of species $\alpha$ and $\beta$, respectively at positions ${\bf r}_{1}$ and ${\bf r}_{2}$and interacting through a pairwise potential which depends only on the distance between the particles,  $u_{\alpha\beta}\equiv u_{\alpha\beta}\left(|{\bf r}_2-{\bf r}_1|\right)$, we choose a coordinate system where ${\bf r}={\bf r}_2-{\bf r}_1$  is the separation vector and  ${\bf R}=w{\bf r}_1+(1-w){\bf r}_2$  the center of motion, with $w\equiv D_{\beta}/(D_{\alpha}+D_{\beta})=\zeta_{\alpha} T_{\beta} /(\zeta_{\alpha} T_{\beta}+\zeta_{\beta} T_{\alpha})$. With this choice, the diffusive motions along the directions ${\bf r}$ and ${\bf R}$ are statistically independent \cite{grosberg2015}. As a result, we obtain the two-particle Langevin dynamics in the form:

\begin{align}\label{eq:2}
\dot{\bf r}=-\frac{1}{\zeta_{r}^{\alpha\beta}}  \partial_{{\bf r}} u_{\alpha\beta}(r) + (2 D_r^{\alpha\beta})^{1/2} {\bm \xi}_{r} (t),\\
\dot{\bf R}=-\frac{1}{\zeta_{R}^{\alpha\beta}} \partial_{{\bf r}} u_{\alpha\beta}(r) + (2 D_R^{\alpha\beta})^{1/2} {\bm \xi}_{R} (t)\label{eq:3}
\end{align}
as $\langle{\bm \xi}_{k}(t)\cdot{\bm 
\xi}_{l}(t')\rangle=d \, \delta(t-t')\delta_{kl}$ and $\langle{\bm \xi}_{k,l}(t)\rangle=0$ where $D_r^{\alpha\beta}=T_{\alpha\beta}/\zeta_r$ and 
$\zeta_r ^{\alpha\beta}=\frac{\zeta_{\alpha}\zeta_{\beta}}{\zeta_{\alpha}+\zeta_{\beta}}$, 
$T_{\alpha\beta}=\frac{\zeta_{\beta}T_{\alpha}+\zeta_{\alpha}T_{\beta}}{\zeta_{\alpha}+\zeta_{\beta}}$, 
$D_R ^{\alpha\beta}=\frac{T_{\alpha} T_{\beta}}{\zeta_{\alpha} T_{\beta}+\zeta_{\beta} T_{\alpha}}$, and
$\zeta_R^{\alpha\beta}=\frac{\zeta_{\beta}T_{\alpha}+\zeta_{\alpha}T_{\beta}}{T_{\alpha}-T_{\beta}}$. The advective term in $\dot{\bf R}$ leads to energy transfer, it only exist for nonequilibrium systems, and it vanishes as $\zeta_R^{-1}\rightarrow 0$ for $T_{\alpha}=T_{\beta}$.

We may now proceed to derive long-time dynamical coefficients for a tagged particle of species $\alpha$ \ef{that has temperature $T_{\alpha}$, friction coefficient $\zeta_{\alpha}$ and diameter $\sigma_{\alpha}$ which plays a role in excluded-volume interactions}. Throughout this paper, we define each secondary particle as type-$\beta$, hence $\{\beta\}$ represents a set of different particle types surrounding the tagged particle. Accordingly, for an $\alpha$-type tagged particle, there are $N_{\beta}$ pairs for $\alpha\beta$ interactions (we take $N_{\alpha}-1\approx N_{\alpha}$), and we leverage the statistical equivalence of pair interactions described in Eqs. \eqref{eq:2} and \eqref{eq:3}.
\begin{figure*}
  \centering \includegraphics[width=2.1\columnwidth]{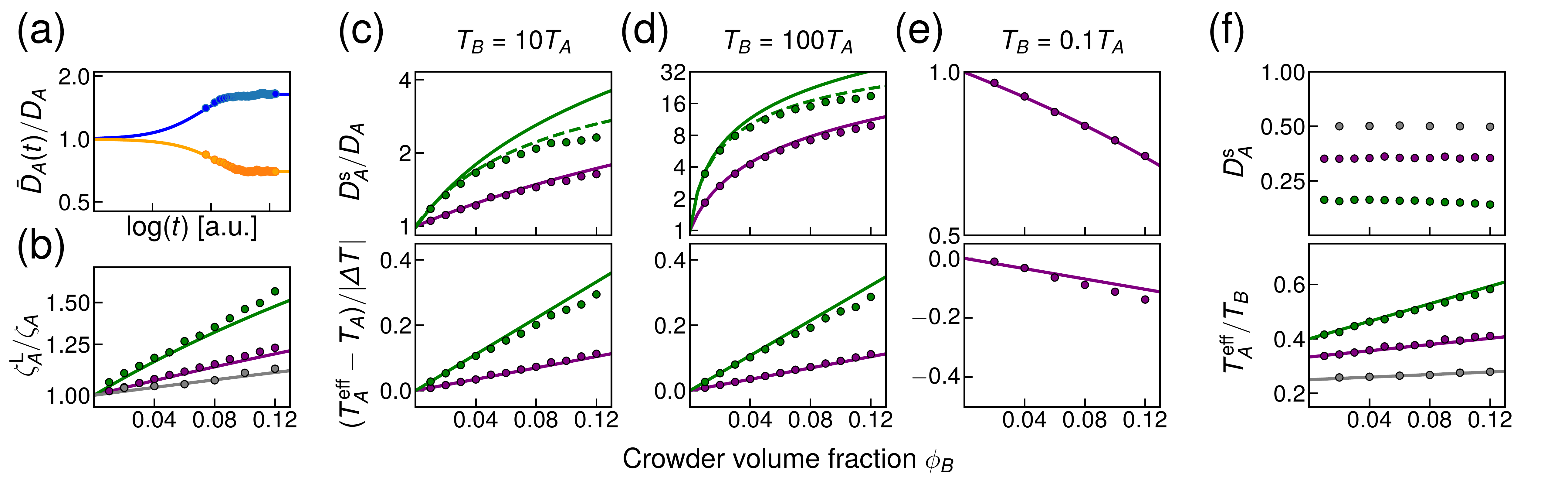}
  \caption{ Tagged particle $A$ in a bath of crowders $B$ in $d=3$. (a) $\bar{D}_{A}(t)$ from simulations (Eq. \eqref{eq:7}, shown in log scale) as a function of time interval $\log(t)$  for two examples; cold particle in hot bath (blue), hot particle in cold bath (orange). In (b)-(f), the color code depicts size ratios $\sigma_A/\sigma_B=0.5,1,2$ in gray, purple and green, respectively, and dots are obtained from simulations, solid lines represent theoretical results for friction coefficient, self-diffusion coefficient, and effective temperature shift respectively from Eqs. \eqref{zetafinal},\eqref{dsfinal},\eqref{eqteff} with scaled contact distances $\sigma_{\alpha\beta}\rightarrow\sigma_{\alpha\beta}'$. In (b), the long-time friction coefficient increases with crowder volume fraction $\phi_B$ and displays a behavior beyond Stokes' law as $\zeta_A ^{\rm L} \sim \sigma_A [1+\epsilon(\phi_B){\sigma_{AB}}^2]$ where $\epsilon(\phi_B)=4\phi_B/{\sigma_B}^2$. Panels (c)-(e) show how both  self-diffusion (shown in log scale) and temperature vary with sizes, temperatures, and volume fraction of crowders. A slight discrepancy between the theory, Eq. \eqref{dsfinal}, and simulations for self-diffusion arises for larger tracer particles for which Eq. \eqref{ddag} (dashed lines) show better success. In (f), we show three illustrative cases where the energy transfer balances frictional forces for temperature ratios satisfying Eq. \eqref{eqstall}.} \label{fig:2}
\end{figure*}

\subsection{Long-time friction coefficient}
At long times, the collisions with the surrounding particles alters the effective friction on the tagged particle. 
In order to determine the density dependent long-time friction 
coefficient, we apply a small constant test force ${\bf F}$ on a tagged  particle of species
$\mathcal{\alpha}$ at
position ${\bf r}_1$ along the lines of Refs. \cite{lekkerkerker1984calculation,van1985exact}. The test force $\bf F$ breaks the symmetry of the steady state and creates a non-homogeneous distribution of the other particles around the tagged one. In return, this alters the force balance and induces an effective mean force ${\bf F}_{\text{in}}$
on the tagged particle through the excluded-volume interactions with the surrounding particles. The average velocity follows a linear relation with the total applied forces $\langle {\bf u}_{\mathcal{\alpha}} 
\rangle=\left({\bf F}+{\bf F_{in}}\right)/\zeta_{\alpha}$ from which we can infer the long-time friction coefficient $\zeta_{\alpha}^{\rm L}$ as we look for $\langle {\bf u}_{\mathcal{\alpha}} 
\rangle={\bf F}/\zeta_{\alpha}^{\rm L}$. To account for all pair interactions with the tagged particle, we define the concentrations for each species $\{\beta\}$ with 
$N_{\beta}$ particles of diameter $\sigma_{\beta}$ in a volume $V$ as $c_{\beta}=N_{\beta}/V$, and the volume fractions as $\phi_{\beta}=\frac{\Omega_d}{2^d d}(\sigma_{\beta})^d c_{\beta}$, where $\Omega_d$ is the solid angle in $d$ dimensions. We obtain  \cite{supplement}
\begin{equation}\label{zetafinal}
\zeta_{\alpha}^{\rm L}=\zeta_{\alpha}\left(1+\sum_{\beta}4\frac{\zeta_r^{\alpha\beta}}{\zeta_{\alpha}}\left(\frac{\sigma_{\alpha\beta}
}{\sigma_{\beta}}\right)^{d}\phi_{\beta}\right)
\end{equation}
in $d=2,3$ where $\sigma_{\alpha\beta}=(\sigma_{\alpha}+\sigma_{\beta})/2$ is the 
contact distance.  The summation is performed over all particle species, and hence our
theory is valid for arbitrary polydisperse compositions with many particle species at 
linear order in concentrations. Including the size effects on solvent-based friction constants \footnote{We use Stokes' values for the friction coefficients while we still neglect hydrodynamic interactions that arises from the flow field.}, this suggests a weak compositional dependence in $d=2$ in accordance with experimental results 
\cite{thorneywork2017self} and a strong dependence in $d=3$ as we illustrate with
simulations below. We should note that the long-time friction coefficients 
are  temperature independent and identical for equilibrium systems of particles in 
contact with a single thermostat and for non-equilibrium systems with the same 
types of particles having multiple temperatures. This calculation is sufficient 
to determine the self-diffusion coefficient at linear order in equilibrium systems where $D^{\rm s} _{\alpha}=D_{\alpha} \zeta_{\alpha}/\zeta_{\alpha}^{\rm L}$. However, the 
non-equilibrium counterpart for the self-diffusion coefficient $D^{\rm s} _{\alpha}$ 
requires an investigation of the dynamics of the tagged-particle density. 

\subsection{Long-time diffusion constant}
The free diffusion constant of an $\alpha$-type particle is given by $D_{\alpha}=T_{\alpha}/\zeta_{\alpha}$ and reflects the behavior at short timescales in the absence of collisions. The long-time diffusion constant is obtained from the asymptotic behavior of the MSD
$\langle  \left(r_{\alpha} (t)-r_{\alpha}(0) \right)^2 \rangle$ of tagged particles as $t\rightarrow \infty$
(averaged over all the $\alpha$ particles from many realizations). We may define 
\begin{equation}
     \bar{D}_{\alpha}(t)=\frac{\langle  \left(r_{\alpha} (t)-r_{\alpha}(0) \right)^2 \rangle}{2d t}.\label{eq:7}
\end{equation}
By definition, $\bar{D}_{\alpha}(0^+)=D_{\alpha}$ and $\bar{D}_{\alpha}(\infty)=D^{\rm s}_{\alpha}$.

The self-diffusion constant $D^{\rm s}_{\alpha}$ is linked to the tagged-particle scattering
function $F({\bf k},t) =\langle e^{-i {\bf k}\cdot \left({\bf r}_1(t)-{\bf r}_1(0) \right)} \rangle$ which is the Fourier transform of the tagged-particle density auto-correlation function. Here, each incident wave is weighted with the steady-state probability 
distribution of finding the tagged particle at ${\bf r}_1(0)$ at $t=0$ and the 
conditional probability of finding the particle at ${\bf r}_1(t)$ at time $t$ given 
that it started at ${\bf r}_1(0)$ (see Supplemental Material\cite{supplement}). For small values of the
Laplace variable $s$ conjugate to $t$, and small wave vector $k$, the Laplace 
transform of the tagged-particle scattering function can be expanded as $s F({\bf 
k},s)\approx1-D_{\alpha}^{\rm s} k^2/s$. The explicit calculation of $D_{\alpha}^{\rm s}$ is 
detailed in Supplemental Material \cite{supplement}.
To first order in concentrations, we obtain
\begin{equation}\label{dsfinal}
D_{\alpha}^s=D_{\alpha}-\sum_{\beta}4\frac{\zeta_r^{\alpha\beta}}{\zeta_{\alpha}}\left(\frac{T_{\alpha}}{\zeta_{\alpha}}+
\frac{T_{\alpha}-T_{\beta}}{\zeta_{\alpha}+\zeta_{\beta}}\right)\left(\frac{\sigma_{\alpha\beta}}{\sigma_{\beta}}\right)^{d}
\phi_{\beta}
\end{equation}
in $d=2,3$ dimensions \footnote{The value of the long-time diffusion constant is negative 
infinity in $d=1$ for hard spheres 
\cite{ackerson1982}}. The second term in the sum reflects the
non-equilibrium contributions to the self-diffusion constant due to the
different temperatures. 

To illustrate certain aspects of Eq. \eqref{dsfinal}, we consider a binary system of particles with one single
particle of type-$A$ (tracer) surrounded by many type-$B$ particles (crowders), i.e., 
$\phi_A\rightarrow 0$ and $\phi_B>0$ while the temperatures differ by an 
amount $\Delta T =T_B-T_A$. When $T_B\gg T_A$ and $\sigma_A\gg \sigma_B$, the fractional change in root-MSD scales linearly with the size of the tracer, i.e., $(\sqrt{D^{\rm s}_A}-\sqrt{D_A})/\sqrt{D_A}\sim \sigma_{A}$, remarkably similar to the observations on tracer diffusion in bacterial cytoplasm where the metabolic activity is tuned by energy depletion procedures \cite{parry2014bacterial}. For equal-sized $A$ and $B$ particles with particle size $\sigma_A=\sigma_B$ and $\zeta_A=\zeta_B=\zeta$, the self-diffusion constant of particle $A$ 
becomes  $D_A^{\rm s}=D_A(1-2\phi_B)+(\Delta T/\zeta) \phi_B$. For equilibrium systems, 
$\Delta T=0$, and the self-diffusion constant recovers the known value 
$D_A^{\rm s}=D_A(1-2\phi_B)$ \cite{hanna1982self,ackerson1982}. For non-equilibrium systems,
as $\Delta T\neq 0$, we investigate two opposite limits: (i) When $T_A\gg T_B$, 
$D_A^{\rm s}\approx D_A(1-3\phi_B)$ increasing the slowing down of the hot particle by the cold 
crowders. This effect is also observed with active Brownian particles for which the 
propulsion speed is more damped \cite{stenhammar2015activity} when the surrounding
bath is composed of less active particles \footnote{In fact, a naive mapping from this
work\cite{stenhammar2015activity}, would suggest $D_A^{\rm s}\approx D_A(1-\beta\phi_B)$ 
with $\beta=2.42$. The reduction is larger than the equilibrium result, i.e., $\beta>2$ but lower than the hard-sphere prediction as $\beta<3$ since the pairwise potential is a soft one (see Eq.\eqref{sigsoft}).}.  (ii) When $T_A\ll T_B$, $D_A^{\rm s}\approx D_B\phi_B$ such 
that the diffusion of cold particles is mainly driven by hot crowders. 

\subsection{Modified Einstein relation and effective temperatures}---We can define an effective temperature $T_{\alpha}^{\text{eff}}$ for the tagged particle  by imposing an Einstein relation 
$D_{\alpha}^{\rm s}= T_{\alpha}^{\text{eff}}/\zeta_{\alpha}^{\rm L}$ (see Supplemental Material \cite{supplement}), and we find
\begin{equation}
T_{\alpha}^{\text{eff}} = T_{\alpha}-\sum_{\beta}4\frac{\zeta_{\alpha}\zeta_{\beta}}{\left(\zeta_{\alpha}+\zeta_{\beta}
\right)^2}(T_{\alpha}-T_{\beta})\left(\frac{\sigma_{\alpha\beta}}{\sigma_{\beta}}\right)^{d}\phi_{\beta}.\label{eqteff}
\end{equation}
Thus, this measure corresponds only to the translational motion and the difference between the effective temperature and the real thermostat temperature reflects the energy exchange between hot and 
cold particles. For binary mixtures of equal-sized hard spheres, this simplifies to $T_{A}^{\text{eff}}=T_{A}+\Delta T\phi_{B}$. 
Since the interaction potential is conservative the total energy conservation imposes that \cite{smrek2017,chubak2020emergence}
\begin{equation}\label{teff}
  \sum_{\alpha} T_{\alpha}c_{\alpha}= \sum_{\alpha} T_{\alpha}^{\text{eff}}c_{\alpha}
\end{equation}
which naturally follows from Eq. \eqref{eqteff}.

\subsection{Brownian dynamics simulations and results in 3 dimensions}
To test our predictions, we perform Brownian dynamics simulations of Eq.~\eqref{eq:1} with a soft repulsive pairwise potential, 
$v_{\alpha\beta}(r)=\frac{k}{2}(\sigma_{\alpha\beta}-r)^2$ for $r<\sigma_{\alpha\beta}$ and $v_{\alpha\beta}(r)=0$ otherwise. 
For a short-range potential with sufficiently narrow cut-off, an effective hard-core interaction diameter can be defined in $d=2,3$, and it reads
\begin{equation}\label{sigsoft}
    \sigma'_{\alpha\beta}=\left(\frac{dB_{\alpha\beta}}{\Omega_d}\right)^{1/d},
\end{equation}
where $B_{\alpha\beta}=\Omega_d\int (1-e^{-v_{\alpha\beta}(r)/T_{\alpha\beta}})r^{d-1}dr$ is the second virial coefficient \cite{grosberg2015,ilker2020phase} which recovers $\sigma'_{\alpha\beta}=(\sigma_{\alpha}+\sigma_{\beta})/2$ for hard spheres. This transformation allows for an equivalent hard-sphere description with scaled contact distances \cite{barker1967,hansen1995,rowlinson1964,andersen1971} using which we can compare the simulation results with our analytical formulas from Eqs.\eqref{zetafinal},\eqref{dsfinal},\eqref{eqteff} by replacing $\sigma_{\alpha\beta}$ 
with $\sigma'_{\alpha\beta}$ (see Supplemental Material for 
details \cite{supplement}).

In Fig.\ref{fig:2}(a), we show $\bar{D}_{\alpha}(t)$ (Eq. \eqref{eq:7}) obtained from simulations and the fitting curve from which we determine the $D_{\alpha}^{\rm s}$ values for two 
examples a tagged hot particle in a cold particle bath (orange) and a tagged cold particle in a hot particle bath (blue). We extract  $D^{\rm s}_{\alpha}$  by extrapolating Eq. \eqref{eq:7} with a stretched exponential, i.e., $\bar{D}_{\alpha}(t)=D_{\alpha}+(D^{\rm s}_{\alpha}-D_{\alpha})\left(1-e^{-(t/\tau)^{\gamma}}\right)$
and typically $\gamma\approx 1/2$ as previously employed in hard-sphere simulations 
\cite{cichocki1990dynamic}. The details on the simulations 
and their analysis are given in Sec.V of the Supplemental Material \cite{supplement}. The size dependence enters in Eqs. \eqref{zetafinal},\eqref{dsfinal}, \eqref{eqteff}, both 
through the contact distance $\sigma_{\alpha\beta}$ and the friction coefficient which follow Stokes' law $\zeta_{\alpha}\sim\sigma_{\alpha}$. We
first consider the behavior of a tagged particle $A$ in a bath of crowders $B$. In all examples, we set $\sigma_B=\sigma$, total volume 
$V=5 \times 10^4 \pi\sigma^3/6$, $\kappa\equiv k\sigma^2/T_{\rm max}=10^2$, where $T_{\rm max}=\max (T_A,T_B)$ . In Fig. \ref{fig:2}(b), we observe the size and 
concentration dependence of the long-time friction coefficient. The increase with radius is stronger than 
the Stokes' law as the concentration of crowders increases. Equation \eqref{zetafinal} suggests that $\zeta_A ^{\rm L} \sim \sigma_A [1+\epsilon(\phi_B){\sigma_{AB}}^2]$ where 
$\epsilon(\phi_B)=4\phi_B/{\sigma_B}^2$. This nonlinear scaling of the friction constant with increasing tracer size is observed in cytoplasmic transport and
polymeric systems \cite{kalwarczyk2011comparative}. Although the interactions in those systems are not limited to steric interactions but also include 
hydrophobic and electrostatic interactions, one can use our formalism extended to narrow-range soft potentials, which defines an effective contact 
distance $\sigma_{\alpha\beta}'$. In Fig. \ref{fig:2}(c)-(e), we show the self-diffusion constant and energy transfer respectively for $T_B=10 T_A$, 
$T_B=100 T_A$, $T_B=0.1 T_A$. Our theory fits particularly well the simulations for the effective temperatures, whereas we observe slight discrepancy in 
self-diffusion constant as the size ratio $\sigma_A/\sigma_B$ increases. 
To propose a correction, we attempted to write
\begin{equation}
{D_{\alpha}^s} ^{\dagger}  = \frac{T_{\alpha}^{\text{eff}}}{\zeta_{\alpha}^{\rm L}}\label{ddag}
\end{equation}
using $T_{\alpha}^{\text{eff}}$ from Eq. \eqref{eqteff}, and $\zeta_{\alpha}^{\rm L}$ from Eq. \eqref{zetafinal}. The outcome is shown by dashed lines and 
and slightly improves the agreement with the simulation results. It obviously recovers the rigorously-derived Eq. \eqref{dsfinal} at linear order, while it
is an uncontrolled expansion at higher orders.

An interesting limit is that where energy transfer balances the friction from the same type of surrounding particles. This happens when
\begin{equation}\label{eqstall}
    \frac{T_A}{T_B}=\frac{1}{2+\sigma_B/\sigma_A},
\end{equation}
then $D_A^{\rm s}=D_A$ becomes insensitive to $\phi_B$, see Fig. \ref{fig:2}(f). Thus, the friction of the $B$ bath vanishes and the crowder particles 
appear completely 
transparent to the tracer particle. The motion remains diffusive (only solvent-based) at all timescales (but strictly $t\gg \tau_{\rm r}$ for systems with active 
swimmers) as the only process is collisions with non-viscous crowders. 

\begin{figure}
  \centering \includegraphics[width=\columnwidth]{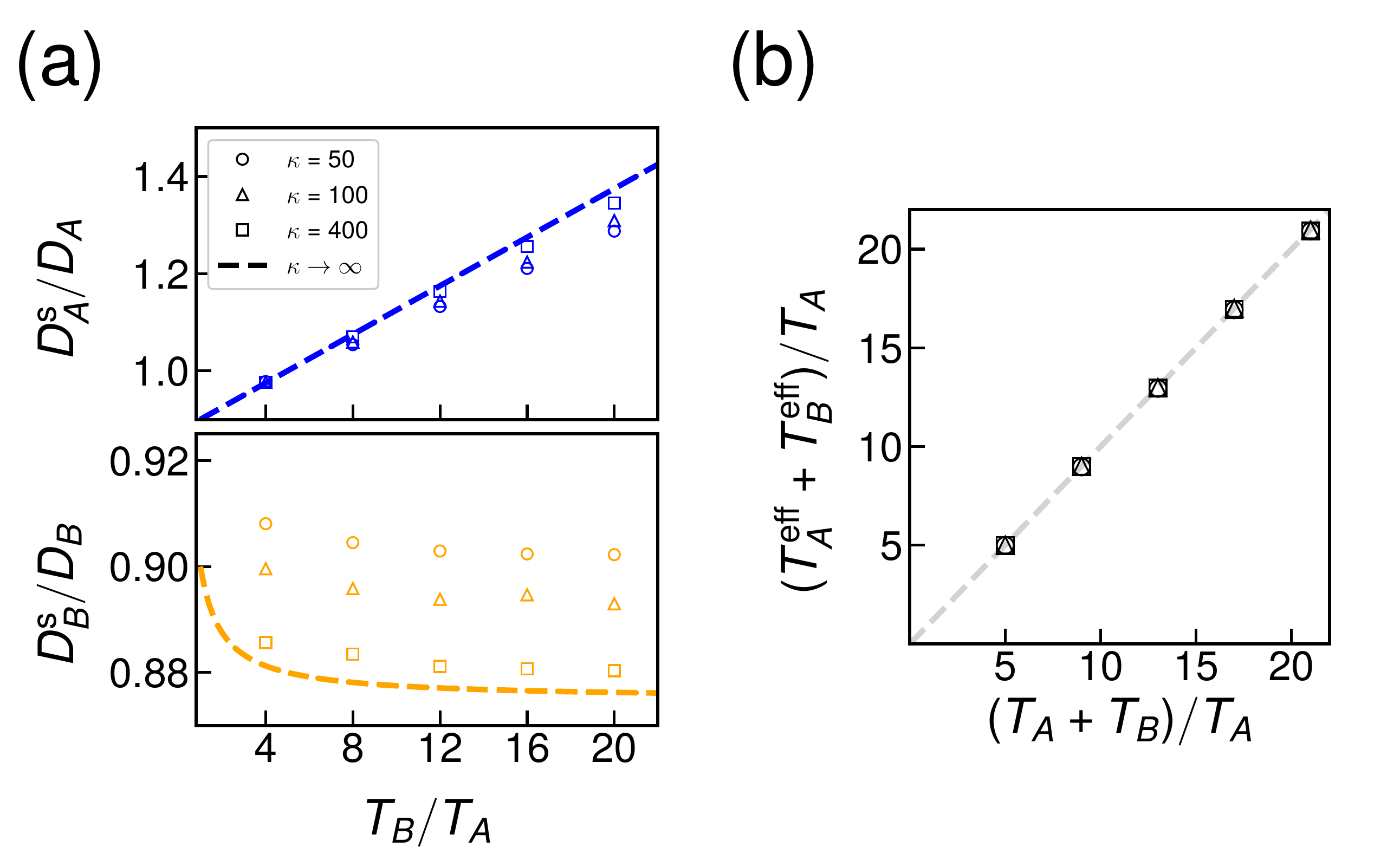}
  \caption{Mixtures, $\kappa$ dependency and energy conservation. In (a), we show the self-diffusion coefficients in free-diffusion coefficient units for equal-sized cold (top) and hot (bottom) particle species in a mixture with $\phi_A=\phi_B=0.025$ for varying values of potential stiffness $\kappa=50,100,400$, respectively shown as circles, triangles and squares. As $\kappa\rightarrow \infty$, the self-diffusion coefficients converge to the hard-sphere values (dashed line). In (b), we display the energy conservation Eq.\eqref{teff} resulting the condition $T_A ^{\rm eff}+T_B ^{\rm eff}=T_A+T_B$ for $c_A=c_B$ which is conserved irrespective of the potential stiffness.}\label{fig:3}
\end{figure}

Next, we study binary mixtures of equal-sized particles at finite concentrations, e.g., $\phi_A=\phi_B=0.025$ and $T_B/T_A>1 $. We also investigate the 
dependence on potential stiffness and the self-diffusion coefficients converge to the hard-sphere values with increasing $\kappa$ as $\lim_{\kappa\rightarrow\infty}\sigma_{\alpha\beta}'=\sigma_{\alpha\beta}$. Upon increasing the temperature ratio, the fractional 
increment in diffusion coefficient of cold particles is significant and follows a linear trend. On the other hand, the fractional loss for hot particles is 
less sensitive to $T_B/T_A$ and converges to $D_B^{\rm s}/D_B=1-3\phi_A-2\phi_B$ as $T_B\gg T_A$ (and $\kappa\rightarrow\infty$) predicted by our theory. We also verified the energy conservation 
\eqref{teff} for the same system (Fig. \ref{fig:3}(b)). This holds irrespective of the potential stiffness and the energy transfer in this example more strictly 
suggests $T_A ^{\rm eff}+T_B ^{\rm eff}=T_A+T_B$ since $c_A=c_B$.

\section{Nature of excluded-volume interactions}\label{sec3}
On a final note, we stress a crucial distinction on how to model excluded-volume interactions in non-equilibrium systems. For real-world systems, the excluded-volume refers to repulsive steric interactions. In equilibrium systems, these can be treated as a boundary condition for the moving particle in which  interparticle distances $r<\sigma_{\alpha\beta}$ are not allowed. This would produce the same statistics for more realistic potentials with nearly hard-sphere descriptions (e.g., Lennard-Jones potential at high temperatures) while both simplifying analytical approaches and allowing for Monte-Carlo methods. However, in such description, the diffusion of a moving particle is always constrained by the existence of the surrounding particles. Thus, even for the colder particle, we would expect no increase in diffusivity, such that $D_{\alpha}^{\rm s}\leq D_{\alpha}$ for all temperature ratios. On the other hand, this would no longer be an obligation for repulsive interactions, and the cold particle can gain energy through collisions. \ef{Conventional Monte-Carlo methods would fail to mimic this effect as jump rates are only constrained ---and not enhanced--- by the existence of another particle. Yet, a proper description can be recovered in lattice models by adding a simple collision rule as demonstrated in Ref.  \cite{abbaspour2021enhanced}.}

Here, our derivation conserves the analytical properties of pair distribution functions, and enables us to approach the hard-sphere limit through a smooth function (and vice versa towards soft potentials). We are therefore able to capture non-equilibrium aspects while keeping the simplicity of a hard-sphere description.

\section{Conclusion}
In this work, we have identified the effect of energy flux and friction on the long-time self-diffusion constant of colloidal particles. We intended to study non-equilibrium mixtures, but Eq. \eqref{dsfinal} also provides an explicit formula that can be used in equilibrium systems of polydisperse solutions at $T_{\alpha}=T_{\beta}=T$ (more generally with $\sigma_{\alpha\beta}\rightarrow \sigma_{\alpha\beta}'$). \ef{We left out the effect of hydrodynamic interactions in our current analysis,  where we focused on the long-time dynamics. However,  hydrodynamic interactions could be included in our framework. Yet this would be a more ambitious undertaking, because also there is no consensus even for equilibrium systems on the significance of the contribution of hydrodynamic interactions to the long-time  value of the self-diffusion coefficient $D_{\alpha}^s$,compared to the value only due to direct collisions  \cite{hanna1982self,batchelor1983diffusion,cichocki1988long,van1992long,zahn1997hydrodynamic}. It would be interesting to investigate these effects on tagged particle dynamics at different time windows in a study complemented by experimental results.}

We hope that our theory will also shed light on experimental studies of active colloids, microswimmers and single-molecule dynamics in biological mixtures such as the cytoplasm. Notably, the theoretical predictions on the size scalings of diffusion enhancement and long-time friction at dilute concentrations of crowders show close resemblance with earlier experimental results on the cytoplasmic transport. Even though we set our theory in the mixing regime, it can be adapted to complex fluids that have spatial heterogeneities by introducing local values of the volume fractions. This would require further study and we leave this to a future work.

\acknowledgements{We thank A.S. Vishen, V. Démery and M. Delarue for interesting discussions. We acknowledge support from the Labex CelTisPhyBio (ANR-11-LABX-0038, ANR-10-IDEX-0001-02) and from Agence Nationale de la Recherche  (ANR), grant ANR-17-CE11-0004}.
\bibliography{mainv1}

\pagebreak
\widetext
\begin{center}
\textbf{\large Supplemental Material for ``Long-time diffusion and energy transfer in polydisperse mixtures of particles with different temperatures''}
\end{center}
\setcounter{equation}{0}
\setcounter{figure}{0}
\setcounter{table}{0}
\setcounter{section}{0}
\setcounter{page}{1}
\makeatletter
\renewcommand{\theequation}{S\arabic{equation}}
\renewcommand{\thefigure}{S\arabic{figure}}

\section{Time evolution of particle distributions}
\subsection{Multi-particle system}
The Langevin dynamics can be reformulated as a Fokker-Planck equation for a multi-particle probability distribution $P(\{{\bf r}\})$ where $\{{\bf r}\}$ is a vector whose components are the positions of all the particles. The time evolution obeys $\partial P /\partial t=\mathcal{L}_{ N}P$ where $\mathcal{L}_{N}$ is the $N$-particle Liouville operator 
\begin{equation}
\mathcal{L}_{N}=\sum_{m=1}^{N}  \frac{\partial}{\partial{\bf r}_m}\cdot D_m \left(\frac{\partial}{\partial{\bf r}_m}+T_m^{-1}\frac{\partial U_{N}}{\partial{\bf r}_m} \right)\label{lvn}
\end{equation}\\
where $D_m=T_m/\zeta_m$ and  $U_{N}=\frac{1}2\sum_{i,j}u(|{\bf r}^{(i)}-{\bf r}^{(j)}|)+\sum_i u_{\rm v}({\bf r}^{(i)})$ is the total potential energy function. As we shall see later on, the general properties of this operator are important to show the statistical equivalence of certain manipulations that we will use. On the other hand, our main mathematical formulation relies mostly on two-particle interactions. Accordingly, next we consider the two-particle system. 
\subsection{Two-particle system}
The time evolution of the two-particle system $P({\bf r}_1,{\bf r}_2)$ is governed by the Liouvilian operator in \eqref{lvn} for $N=2$. In the main text, we have written the two-particle Langevin dynamics transformed into relative and center of motion coordinates. We follow the same analysis and distinguish particle types $
\alpha$ and $\beta$ that have temperature $T_{\alpha}$ and $T_{\beta}$, friction coefficient $\zeta_{\alpha}$ and $\zeta_{\beta}$ , diameter $\sigma_{\alpha}$ and $\sigma_{\beta}$. We use the coordinates ${\bf r}={\bf 
r}_2-{\bf r}_1$, the separation vector and  ${\bf 
R}=w{\bf r}_1+(1-w){\bf r}_2$ the 
center of motion with $w\equiv D_{\beta}/(D_{\alpha}+D_{\beta})=\zeta_{\alpha} 
T_{\beta} /(\zeta_{\alpha} T_{\beta}+\zeta_{\beta} 
T_{\alpha})$. The corresponding Fokker-Planck equation reads:

\begin{eqnarray}\label{FPrel}
\frac{\partial P_{\alpha\beta}}{\partial t}= \mathcal{L}_{ r} P_{\alpha\beta} +\mathcal{L}_{ R}P_{\alpha\beta}
\end{eqnarray}\\
where the Liouvillian operators are given by:
\begin{eqnarray}\label{lr}
\mathcal{L}_{ r}&=&\frac{\partial}{\partial {\bf r}} \cdot \left[ D_r ^{\alpha\beta} \frac{\partial }{\partial {\bf r}}+\frac{1}{\zeta_r ^{\alpha\beta}} \frac{\partial u_{\alpha\beta}}{\partial {\bf r}} \right],\\\nonumber\\
\mathcal{L}_{R}&=&\frac{\partial}{\partial {\bf R}} \cdot \left[ D_R ^{\alpha\beta}  \frac{\partial }{\partial {\bf R}} +\frac{1}{\zeta_R^{\alpha\beta}} \frac{\partial u_{\alpha\beta}}{\partial {\bf r}} \right]\label{lrr}
\end{eqnarray}
with $D_r^{\alpha\beta}=T_{\alpha\beta}/\zeta_r$ and $\zeta_r ^{\alpha\beta}=\frac{\zeta_{\alpha}\zeta_{\beta}}{\zeta_{\alpha}+\zeta_{\beta}}$, $T_{\alpha\beta}=\frac{\zeta_{\beta}T_{\alpha}+\zeta_{\alpha}T_{\beta}}{\zeta_{\alpha}+\zeta_{\beta}}$, $D_R ^{\alpha\beta}=\frac{T_{\alpha} T_{\beta}}{\zeta_{\alpha} T_{\beta}+\zeta_{\beta} T_{\alpha}}$, $\zeta_R^{\alpha\beta}=\frac{\zeta_{\beta}T_{\alpha}+\zeta_{\alpha}T_{\beta}}{T_{\alpha}-T_{\beta}}$. 

The steady-state solution of the Fokker Planck equation 
obtained by separation of variables, 
$P_{\alpha\beta}^{\text{ss}}\equiv G_{\alpha\beta}^{\text{ss}}({\bf R}) g_{\alpha\beta}^{\text{ss}}({\bf r})/V^2$ reads:
\begin{eqnarray}\label{eqgss}
g_{\alpha\beta} ^{\text{ss}}(r)= \exp [ -u_{\alpha\beta} (r)/T_{\alpha\beta}]
\end{eqnarray}\\
and $G^{\rm ss}_{\alpha\beta}({\bf R})$ is uniform. Note that this sets the flux along ${\bf r}$ to zero while a non-zero flux would exist along ${\bf R}$ direction reflecting the non-equilibrium nature of the problem.
\section{Necessary identities and relations}\label{sec1}
We now introduce the mathematical identities, which are required to treat the pairwise hardcore interaction terms in our derivation.  
The steady-state pair distribution function for the two-body problem as given above reads

\begin{eqnarray}\label{eqgss1}
g_{\alpha\beta} ^{\text{ss}} (r) &= e^{ -u_{\alpha\beta} (r)/T_{\alpha\beta}}\\
& =  \Theta(r-\sigma_{\alpha\beta})\nonumber
&
\end{eqnarray}
where the second equality holds for a hard-sphere potential only for which it can be expressed in terms of a Heaviside step function where $\sigma_{\alpha\beta}=(\sigma_{\alpha}+\sigma_{\beta})/2$ is the contact distance between the corresponding two particles. 
The spatial derivative of the distribution function is then given by a Dirac delta function: 
\begin{eqnarray}\label{eq3}
\frac{\partial}{\partial r} e^{ -u_{\alpha\beta} (r)/T_{\alpha\beta}}=-\frac{e^{ -u_{\alpha\beta} (r)/T_{\alpha\beta}}}{T_{\alpha\beta}}\frac{\partial u_{\alpha\beta}(r)}{\partial r}=\delta(r-\sigma_{\alpha\beta}).
\end{eqnarray}\\
These  identities  will  allow  us  to  keep  track  of  pairwise  quantities  and  to  simplify  the  solution,  because the  Dirac  delta  function  can be treated as a boundary condition as we will demonstrate below in our calculations.

\section{Long-time dynamic behavior of a tagged particle}
Here, we detail the steps to calculate the long-time dynamical constants of a tagged particle of type $\alpha$ in the presence of surrounding $\{\beta\}$ particles. As mentioned in the main text, the first-order corrections in particle densities emerge from summing over all two-particle clusters. Thus, once we solve the two-body problem, we can simply sum over all clusters that the tagged particle forms with the surrounding $\{\beta\}$ particles. 
\subsection{Long-time viscosity}\label{sec2a}
 At long times, the long-time friction constant on the tagged particle is a function of the volume fractions of the surrounding particles, i.e.,  $\zeta_{\alpha}^{\rm L}\equiv \zeta_{\alpha}^{\rm L}(\{\phi_{\beta}\})$. In order to determine the 
density-dependent friction constant, we apply a small constant test force ${\bf F}$ on a single particle of type $\mathcal{\alpha}$ 
at position ${\bf r}_1$ (the tagged particle) along the lines of Refs. 
\cite{lekkerkerker1984calculation,van1985exact}. In the presence of the external test force, the modified Fokker-Planck equation for the pair distribution 
function is
\begin{equation}
\frac{\partial g_{\alpha\beta} ({\bf r})}{\partial t}= D_r \nabla_{\bf r}\cdot \left[\nabla_{\bf r} g_{\alpha\beta} ({\bf r}) + \frac{1}{T_{\alpha\beta}} g_{\alpha\beta} ({\bf r}) \nabla_{\bf r} u_{\alpha\beta} (r) \right]+\frac{{\bf F}}{\zeta_{\alpha}}\cdot \nabla_{\bf r} g_{\alpha\beta} ({\bf r}).
\end{equation}
The steady-state solution of this equation can be obtained by introducing a perturbation to the solution of Eq.\eqref{eqgss}. This suggests that up to linear order in ${\bf F}$, we have
\begin{equation}\label{gmod}
g_{\alpha\beta} ^{\text{ss}'}  ({\bf r},{\bf F})=e^{ -u_{\alpha\beta} (r)/T_{\alpha\beta}}\left[ 1+\frac{q(r)}{r}{\bf r}\cdot {\bf F}\right],
\end{equation}
where $q(r)$ is a function of the distance $r$ only. As a result, the equation for the steady-state pair distribution function reduces to:
\begin{equation}\label{eqlong}
D_r \nabla_{\bf r}\cdot \left[e^{ -u_{\alpha\beta} (r)/T_{\alpha\beta}}\nabla_{\bf r}  \frac{q(r)}{r}{\bf r}\cdot {\bf F} \right]+\frac{{\bf F}}{\zeta_{\alpha}}\cdot \nabla_{\bf r} e^{ -u_{\alpha\beta} (r)/T_{\alpha\beta}}=0.
\end{equation}\\
Then, in $d=2,3$, the radial-dependent part $q(r)$ satisfies the relation
\begin{equation}
e^{ -u_{\alpha\beta} (r)/T_{\alpha\beta}}
\left[\frac{d}{dr}
\left(r^{d-1}\frac{dq}{dr}\right)-(d-1)r^{d-3}q\right]+\delta (r-\sigma_{\alpha\beta})r^{d-1}\left(\frac{dq}{dr}+\frac{1}{\zeta_{\alpha}D_r^{\alpha\beta}}\right)=0, \label{eqlong2}
\end{equation}
where we used Eq.~\eqref{eq3}. We first look for a solution in the range where $r>\sigma_{\alpha\beta}$ and then use the part that is coupled with the Dirac delta function as a boundary condition at $r=\sigma_{\alpha\beta}$ such that we satisfy Eq.\eqref{eqlong2} everywhere. Thus, we look for a solution of the following equation:
\begin{equation}\label{dqdr0}
\frac{d}{dr}
\left(r^{d-1}\frac{dq}{dr}\right)-(d-1)r^{d-3}q=0
\end{equation}
with boundary conditions
\begin{subequations}\label{qbcs}
\begin{eqnarray}\label{boundary_condition1}
    \frac{dq}{dr}\Big |_{r=\sigma_{\alpha\beta}}&=&-\frac{1}{\zeta_{\alpha} D_r ^{\alpha\beta}},\\\label{boundary_condition2}
    q(\infty)&=&0.
\end{eqnarray}
\end{subequations}
The boundary condition \eqref{boundary_condition2}  is necessary to obtain a valid solution in a large volume $V$. The solution of Eq. \eqref{dqdr0} is of the form $q(r)=A_1 r^{1-d}+A_2 r$, and the condition $q(\infty)=0$ implies that $A_2=0$. By applying the other boundary condition \eqref{boundary_condition1}, we obtain:
\begin{equation}\label{qfin}
q(r)=\frac{{\sigma_{\alpha\beta}}^d}{(d-1)\zeta_{\alpha}D_r ^{\alpha\beta}}r^{1-d}
\end{equation}
which can then be plugged in to obtain the full form of the modified steady-state pair distribution function Eq.\eqref{gmod}.

For the $N$-particle system with $N=\sum_{\beta} N_{\beta}$, the modified steady-state pair distribution function (Eq.\eqref{gmod}) induces a relaxation force on the tagged particle exerted by $N-1$ particles. By summing over $N-1$ pairs, and accounting for different particle species, we obtain:
\begin{equation}
{\bf F}_{\text{in}}=\sum_{\beta} c_{\beta}\int g_{\alpha\beta} ^{\text{ss}'}  ({\bf r},{\bf F}) \frac{ d u_{\alpha\beta} (r)}{dr}\hat{\bf r}{d {\bf r}}
\end{equation}
which is a summation over all secondary particle types (including $\alpha$) where $\hat{\bf r}={\bf r}/r$, $c_{\beta}=N_{\beta}/V$ are the particle concentrations \footnote{There are $N_{\beta}$ pairs for $\alpha\neq\beta$ interactions, $N_{\alpha}-1$ pairs for  interactions with its own species and we take $N_{\alpha}-1\approx N_{\alpha}$.}.
The integral is in  $d$ dimensions with a volume element $d{\bf r}=r^{d-1}dr d\Omega_d$ where $\Omega_d$ is the solid angle in $d$ dimensions. Next, we use spherical symmetry $\int \hat{\bf r} d\Omega_d=0$ and $\int ({\bf F}\cdot \hat{\bf r})\hat{\bf r}d\Omega_d={\bf F}\Omega_{d}/d$ and plug in the solution \eqref{gmod} along with Eq.\eqref{qfin}. As a result, we get 
\begin{equation}
    {\bf F}_{\text{in}}=-\sum_{\beta}4\frac{\zeta_r}{\zeta_{\alpha}}\left(\frac{\sigma_{\alpha\beta}}{\sigma_{\beta}}\right)^{d}\phi_{\beta} {\bf F}
\end{equation} 
where the volume fraction is $\phi_{\beta}=\left(\frac{\Omega_d}{2^d d}\right){\sigma_{\beta}}^d c_{\beta}$. The long-time friction constant is defined by the relation between average velocity and force, $\langle {\bf u}_{\mathcal{\alpha}} \rangle=\left({\bf F}+{\bf F_{\text{in}}}\right)/\zeta_{\alpha}={\bf F}/\zeta_{\alpha}^{\rm L}$, leading to 
\begin{equation}\label{zetafinals}
\zeta_{\alpha}^{\rm L}=\zeta_{\alpha}\left[1+\sum_{\beta}4\frac{\zeta_r^{\alpha\beta}}{\zeta_{\alpha}}\left(\frac{\sigma_{\alpha\beta}}{\sigma_{\beta}}\right)^{d}\phi_{\beta}\right]
\end{equation}
which is Eq.(4) in the main text.
\subsection{Self-diffusion coefficient}\label{section2b}
As discussed in the main text, the self-diffusion coefficient can be obtained from the tagged-particle scattering function. 
Using the time-evolution operator, the tagged-particle scattering function can be written as:
\begin{equation}\label{eq13}
F({\bf k},t)=\langle e^{-i {\bf k}\cdot {\bf r}_1} e^{\mathcal{L}_{\text{N}}t} e^{i {\bf k}\cdot {\bf r}_1}\rangle_{\text{ss}},
\end{equation} 
where the brackets $\langle \rangle_{\text{ss}}$ denotes right averaging, i.e., $\langle f \rangle_{\text{ss}} = \int f P^{\text{ss}}_{\text{N}} d{\bf X}$, $d{\bf X}\equiv d{\bf r}_1d{\bf r}_{2}...d{\bf r}_{N}$, $P^{\text{ss}}_{\text{N}}$ is the multi-particle steady-state distribution function, and any operator inside $f$ acts also on $P^{\text{ss}}_{\text{N}}$ (see Appendix \ref{appendixa}) . For small $s$ and $k$ the Laplace transform of the tagged-particle scattering function  becomes
\begin{equation}
sF({\bf k},s)\approx 1-\frac{D_{\alpha}^{\rm s} k^2}{s}.
\end{equation}
We next use $F({\bf k},s)=\langle e^{-i {\bf k}\cdot {\bf r}_1} \left(s-\mathcal{L}_{\rm N}\right)^{-1} e^{i {\bf k}\cdot {\bf r}_1}\rangle_{\rm ss}$, and isolate the self-diffusion coefficient:
\begin{equation}
D_{\alpha}^{\rm s}=\lim_{s\to 0}\lim_{k\to 0}\left\{\frac{s}{k^2}\langle e^{-i {\bf k}\cdot {\bf r}_1} \left[1-s\left(s-\mathcal{L}_{\rm N}\right)^{-1}\right] e^{i {\bf k}\cdot {\bf r}_1}\rangle_{\rm ss}\right\}.
\end{equation} 
By rearranging the terms, we have the following relation \cite{hanna1982self}:
\begin{equation}\label{sdeq}
D_{\alpha}^{\rm s}=\lim_{s\to 0}\lim_{k\to 0}\left\{-\frac{1}{k^2}\left( \langle e^{-i{\bf k}\cdot  {\bf r_1}} \mathcal{L}_{\rm N}  e^{i{\bf k}\cdot  {\bf r_1}}\rangle_{\rm ss} +  \langle e^{-i{\bf k}\cdot  {\bf r_1}} \mathcal{L}_{\rm N}\left(s-\mathcal{L}_{\rm N} \right)^{-1} \mathcal{L}_{\rm N}  e^{i{\bf k}\cdot  {\bf r_1}}\rangle_{\rm ss}  \right)\right\}.
\end{equation}\\
The first term in  Eq. (\ref{sdeq}) recovers the bare diffusion constant in the absence of interactions between the particles. Applying the operator $\mathcal{L}_{\rm N}$ on the right of the resolvent operator $\left(s-\mathcal{L}_{\rm N} \right)^{-1}$ and integrating by parts from the left side we obtain, in the limit $k\rightarrow0$,
\begin{equation}\label{sdeq1}
D_{\alpha}^{\rm s}=D_{\alpha}-\lim_{s\to 0} \int \frac{1}{\zeta_{\alpha}}\frac{\partial U_{\rm N}}{\partial {\bf r}_1}\cdot{\hat{\bf  k}} \left(s-\mathcal{L}_{\rm N} \right)^{-1}{\hat{\bf  k}}\cdot\left(2D_{\alpha}\frac{\partial P^{\rm ss}_{\rm N}}{\partial {\bf r}_1}+\frac{1}{\zeta_{\alpha}} P^{\rm ss}_{\rm N} \frac{\partial U_{\rm N}}{\partial {\bf r}_1}\right) d{\bf X}.
\end{equation}\\
In a system of $N$ particles, the effect of the interactions must be summed over all the particles. At lowest order in particle densities, we may consider only the contribution from all two-body clusters. Then the two-particle ensemble average of a function is  $\langle f \rangle_{\rm ss}= \frac{1}{V^2}\int \int f g_{\alpha\beta}^{\rm ss}({\bf r})G_{\alpha\beta}^{\rm ss}({\bf R})d{\bf r} d{\bf R}$ and we consider all $\alpha\beta$ pairs identical. Using ${\bf r}_1={\bf R}-(1-w){\bf r}$ for the tagged particle and its differential form, we obtain:

\begin{equation}\label{sdeqr}
D_{\alpha} ^{\rm s}=D_{\alpha}-\frac{1}{\zeta_{\alpha}^2}\sum_{\beta}\left[1+2\left(\frac{T_{\alpha}}{T_{\alpha\beta}}-1\right)\right]\mathcal{I}_{\alpha\beta}c_{\beta},
\end{equation}
where 
\begin{equation}
    \mathcal{I}_{\alpha\beta}=\lim_{s\to 0}\left\langle (\hat{\bf k}\cdot \hat{\bf r})\frac{\partial u_{\alpha\beta}}{\partial { r}} \left(s-\mathcal{L}_{\alpha\beta} \right)^{-1} \frac{\partial u_{\alpha\beta}}  {\partial {r}}  (\hat{\bf k}\cdot \hat{\bf r}) \right\rangle_{\rm ss},
\end{equation}\\
$\hat{\bf x}={\bf x}/x$ is a unit vector parallel to ${\bf x}$, and we performed the summation over $N-1$ pairs composed of a tagged particle of type $\alpha$ and the surrounding $\{\beta\}$ particles. The pairwise Liouvillian operator is given by $\mathcal{L}_{\alpha\beta}=\mathcal{L}_{r}+\mathcal{L}_{R}$ using definitions from Eqs.\eqref{FPrel}-\eqref{lrr}. Since $G_{\alpha\beta}^{\rm ss}$ is uniform, we may simply integrate over ${\bf R}$, and write  $\langle f \rangle_{\rm ss}= \frac{1}{V}\int f g_{\alpha\beta}^{\rm ss}(r)d{\bf r}$. Then, by using
\begin{equation}g_{\alpha\beta}(r)\frac{\partial u}{\partial r} \hat{\bf r} \cdot \hat{\bf k}=-T_{\alpha\beta}\hat{\bf k}\cdot\nabla_{\bf r}g_{\alpha\beta}(r),
\end{equation}
we get
\begin{equation}\label{iab1}
\mathcal{I}_{\alpha\beta}=-T_{\alpha\beta}\lim_{s\to 0} \int (\hat{\bf k}\cdot \hat{\bf r})\frac{\partial u_{\alpha\beta}}{\partial { r}} \left(s-\mathcal{L}_r \right)^{-1} \chi({\bf r},t=0) d{\bf r}.
\end{equation}
where we replaced the Liouvillian operator as  $\mathcal{L}_{\alpha\beta}\rightarrow \mathcal{L}_r$ since there is no longer any ${\bf R}$ dependence. We also defined a function $\chi({\bf r},t=0)\equiv\hat{\bf k}\cdot\nabla_{\bf r}g(r)$. We can then use the definition of Laplace transforms to write $(s-\mathcal{L}_r)^{-1}\chi({\bf r},t=0)= \int e^{\mathcal{L}_r t}\chi({\bf r},t=0) e^{-st} dt $ and identify the time-evolution operator as $e^{\mathcal{L}_r t}$ to write $\chi({\bf r},t)=e^{\mathcal{L}_r t}\chi({\bf r},t=0)$. Now, we take $s\rightarrow 0$ and get:
\begin{equation}\label{lims0}
    \lim_{s\to 0}(s-\mathcal{L}_r)^{-1}\chi({\bf r},t=0)=\tilde{\chi}({\bf r},s=0)
\end{equation}
where $\tilde{\chi}({\bf r},s=0)$ is the Laplace transform of  $\chi({\bf r},t)$ at $s=0$. 
As $\mathcal{L}_r$ is independent of time, the time evolution imposes that the function $\chi$ is a solution of $\partial \chi({\bf r},t)/\partial t=\mathcal{L}_r\chi({\bf r},t)$. Taking the Laplace transform of this equation gives:
\begin{equation}
    \mathcal{L}_r\tilde{\chi}({\bf r},s)=s\tilde{\chi}({\bf r},s)-\hat{\bf k}\cdot\nabla_{\bf r}g_{\alpha\beta}(r).
\end{equation}
Then, we set $s=0$ and make the ansatz  
\begin{equation}
\tilde{\chi}({\bf r},s=0)=e^{-u(r)/T_{\alpha\beta}}X(r) (\hat{\bf k}\cdot \hat{\bf r})
\end{equation}
which leads to
\begin{equation}
D_r \nabla_{\bf r}\cdot \left[e^{ -u_{\alpha\beta} (r)/T_{\alpha\beta}}\nabla_{\bf r}  \frac{X(r)}{r}{\bf r}\cdot \hat{\bf k} \right]+\hat{\bf k}\cdot \nabla_{\bf r} e^{ -u_{\alpha\beta} (r)/T_{\alpha\beta}}=0.
\end{equation}\\
This equation is equivalent to \eqref{eqlong} with $q(r)\rightarrow X(r)/\zeta_{\alpha}$ and ${\bf F}\rightarrow \hat{\bf k}$. Thus, adapting the solution \eqref{qfin} for $X(r)$, we obtain:
\begin{equation}
    \tilde{\chi}({\bf r},s=0)=e^{-u(r)/T_{\alpha\beta}}\frac{{\sigma_{\alpha\beta}}^d}{(d-1)D_r ^{\alpha\beta}}r^{1-d} (\hat{\bf k}\cdot \hat{\bf r}).
\end{equation}\\
By plugging this result into \eqref{iab1} and using \eqref{lims0}, and the relations
 $d{\bf r}=r^{d-1}drd\Omega_d$, $\int (\hat{\bf k}\cdot \hat{\bf r})^2 d\Omega_d=\Omega_d/d$ and \eqref{eq3}, we get:
\begin{equation}
    \mathcal{I}_{\alpha\beta}=\frac{\Omega_d}{d(d-1)}T_{\alpha\beta}\zeta_r^{\alpha\beta}{\sigma_{\alpha\beta}} ^{d}.
\end{equation}
Finally, by inserting this result in \eqref{sdeqr}, we obtain in $d=2,3$
\begin{equation}\label{dsd3}
    D_{\alpha}^{\rm s}=D_{\alpha}-\sum_{\beta} 4\frac{\zeta_r^{\alpha\beta}}{\zeta_{\alpha}}\left(\frac{T_{\alpha}}{\zeta_{\alpha}}+\frac{T_{\alpha}-T_{\beta}}{\zeta_{\alpha}+\zeta_{\beta}}\right)\left(\frac{\sigma_{\alpha\beta}}{\sigma_{\beta}}\right)^{d}\phi_{\beta},
\end{equation}
where again we used the definition of volume fractions $\phi_{\beta}=\left(\frac{\Omega_d}{2^d d}\right){\sigma_{\beta}}^d c_{\beta}$.

Note that in previous works \cite{hanna1982self,ackerson1982}, in order to get an expression for $\chi({\bf r},t)$, the time evolution equation for a pair of equal-sized hard spheres is introduced by relating it to a conditional probability, i.e., $\chi({\bf r},t)\sim p({\bf r},t| {\bf r}(0),0)$ with the appropriate projection. Then, the solution of the time-evolution equation is expressed as an eigenfunction expansion of the Liouvillian operator which then inserted to replace this part. The final result is obtained in the limit $s\rightarrow 0$. However, as we are only interested in the long-time behavior, this procedure is not necessary and here instead, we took a shortcut which does not require neither conditional probabilities nor an eigen-function expansion, and we used a procedure very similar to the calculation of the long-time friction constant presented in Section \ref{sec2a}.

\subsection{Modified Einstein relation} \label{modifiedeinstein}
Using the results for long-time friction and self-diffusion coefficients, we can model an effective long timescale dynamics for the tagged particle as if it were an isolated particle inside a bath at an effective temperature $T^{\text{eff}}_{\alpha}$.  
We then apply an external force ${\bf F}$ only on the tagged particle, in a similar manner shown in Section \ref{sec2a}. Experimentally, this can be done for example, with an electric field or a laser trap on the particle.  At long enough times, the self-diffusion description suggests that for a particle of type $\alpha$ starting at position ${\bf r}_{1}(0)$ at $t=0$, the conditional probability distribution to find the same particle at ${\bf r}_{1}$ when $\tau$ will obeys an effective dynamics of the form:
\begin{equation}
    \frac{\partial}{\partial t} p_{\alpha}({\bf r}_1,\tau | {\bf r}_{1}(0), 0)=D_{\alpha} ^{\rm s} \nabla_{{\bf r}_1}^2 p_{\alpha}({\bf r}_1,\tau | {\bf r}_{1}(0), 0)-\frac{1}{\zeta_{\alpha}^{\rm L}}  \nabla_{{\bf r}_1} \cdot {\bf F} p_{\alpha}({\bf r}_1,\tau | {\bf r}_{1}(0), 0).
\end{equation}\\
If we define a potential $U_{\alpha}=-\int ^{{\bf r}_1} {\bf F}\cdot {\bf dr}$, the steady-state distribution of this effective dynamics reaches a Boltzmann form $p_{\alpha}({\bf r}_1,\infty | {\bf r}_{0}, 0)\sim e^{-U_{\alpha}/(D_{\alpha} ^{\rm s} \zeta_{\alpha}^{\rm L})}$ suggesting that the effective temperature $T^{\text{eff}}_{\alpha}=D_{\alpha} ^{\rm s} \zeta_{\alpha}^{\rm L}$ which gives the modified Einstein relation in the main text.

Note that for this example, the effective temperature is not necessarily a true bath temperature, but it contributes only on the translational motion of the tagged particle. It basically sets the noise amplitude of the tagged particle for longtime jumps where this effective description is valid. This definition is different than the effective temperature obtained from the velocity distributions \cite{grosberg2018,demery2019driven}.

\section{Interaction volume of soft repulsive forces}
In our analytical approach, we considered particles interacting only through a hard-sphere potential. This  allowed us to leverage 
the identities  \eqref{eqgss1} and \eqref{eq3}, so as to obtain Eqs. \eqref{dqdr0} and \eqref{qbcs}, and to use a similar treatment in Section \ref{section2b}. However, our procedure  does not lead to exact analytical results for a more realistic soft potential $v(r)$ for which the pair distribution function $g_{\alpha\beta}^{\text{ss}}(r)=e^{-v(r)/T_{\alpha\beta}}$ is not a Heaviside step function. On the other hand, we can approximate the effect of this soft potential following a similar approach to the ones by Rowlinson \cite{rowlinson1964}, Barker-Henderson \cite{barker1967}, and Andersen-Weeks-Chandler \cite{andersen1971}.  These approaches originally aimed at mapping the equation of state for a colloidal system with short-range interactions, to an effective hard-sphere description. Here, we propose to apply this approach to long-time dynamical properties by modifying the value of $\sigma_{\alpha\beta}$.

Accordingly, we define the two Mayer functions $f_{\alpha\beta}(r)=1-e^{-v(r)/T_{\alpha\beta}}$ and $f_{\alpha\beta} ^{\text{h}}(r,\sigma_{\alpha\beta})=1-e^{-u(r)/T_{\alpha\beta}}$ where $u(r)$ is hard-sphere potential with contact distance $\sigma_{\alpha\beta}$ to be determined. A macroscopic scalar observable of the many-body system can be written as a functional of $f_{\alpha\beta}(r)$ as 
\begin{equation}
\mathcal{Y}([f_{\alpha\beta}],\Phi)=\int y(f_{\alpha\beta},\nabla f_{\alpha\beta},...,\Phi)d{\bf r},
\end{equation}
where $\Phi=(c_{\alpha},c_{\beta},\sigma_{\alpha},\sigma_{\beta},T_{\alpha},T_{\beta})$ is a set of additional ${\bf r}$-independent parameters. We expand this functional around the hard-sphere value $f_{\alpha\beta}(r)=f_{\alpha\beta} ^{\text{h}}(r,\sigma_{\alpha\beta})$ and obtain
\begin{equation}
  \mathcal{Y}([f_{\alpha\beta}],\Phi)\approx \mathcal{Y}([f_{\alpha\beta}^{\text{h}}],\Phi)+\int \frac{\delta \mathcal{Y}}{\delta f_{\alpha\beta}}\Big |_{f_{\alpha\beta}=f_{\alpha\beta}^{\text{h}}}(f_{\alpha\beta}-f_{\alpha\beta}^{\text{h}})d{\bf r},
\end{equation}
where we neglected higher-order terms. By a proper choice of the contact distance $\sigma_{\alpha\beta}=\sigma'_{\alpha\beta}$, a valid approximation for $\mathcal{Y}([f_{\alpha\beta}],\Phi)\approx \mathcal{Y}([f_{\alpha\beta}^{\rm h}(r,\sigma'_{\alpha\beta})],\Phi)$ can be obtained. One such choice would be to look for the contact radius $\sigma'_{\alpha\beta}$ that satisfies
\begin{equation}
    \int \left[f_{\alpha\beta}(r)-f_{\alpha\beta}^{\text{h}}(r,\sigma'_{\alpha\beta})\right]d{\bf r}=0.
\end{equation}
This means that the effective interaction volume in $d=2,3$ dimensions is given by
\begin{equation}
    \frac{\Omega_d}{d}{\sigma'_{\alpha\beta}} ^d=\int \left( 1-e^{-v(r)/T_{\alpha\beta}} \right) d{\bf r}
\end{equation}
which is equivalent to the expression that we refer to in the main text.
As a matter of fact, such a choice of $\sigma'_{\alpha\beta}$ exactly accounts for the second virial coefficient in the equation of state of soft particles. 

Finally, at this order of the expansion with two-body clusters, we can check the validity of the approximation in terms of softness. The softness parameter $\xi$  is defined as the range $|r/\sigma'_{\alpha\beta}-1|\leq \xi$ in which $f_{\alpha\beta}(r)-f_{\alpha\beta}^{\text{h}}(r,\sigma'_{\alpha\beta})$ is nonzero. Then, it follows that 
\begin{equation}
    \mathcal{Y}([f_{\alpha\beta}],\Phi)=\mathcal{Y}([f_{\alpha\beta}^h(r,\sigma'_{\alpha\beta}],\Phi)\left(1+\mathcal{O}(\xi^2) \right).
\end{equation}
Thus, accordingly, we expect \begin{equation}
D^{\rm s} _{\alpha}([f_{\alpha\beta}],\Phi)=D^{\rm s} _{\alpha}(\sigma' _{\alpha\beta})\left(1+\mathcal{O}(\xi^2) \right),
\end{equation} 
where $D^{\rm s} _{\alpha}(\sigma' _{\alpha\beta})$ is obtained by replacing $\sigma_{\alpha\beta}$ with $\sigma' _{\alpha\beta}$ in Eq. \eqref{dsd3}.

\section{Brownian-Dynamics Simulations}\label{secs5}
We perform Brownian dynamics simulations of a many-body system using Eq.(1) in main text with a soft repulsive pairwise potential of the form,
\begin{equation}
    v_{\alpha\beta}(r)=\frac{k}{2}(\sigma_{\alpha\beta}-r)^2
\end{equation}
for $r<\sigma_{\alpha\beta}$ and $v_{\alpha\beta}(r)=0$ otherwise. In Ref. \cite{weber2016}, certain aspects of these simulations for the same model (particles with different temperatures) have been tested in the context of phase separation. In all examples, we set $\sigma_B=\sigma$, total volume $V=50,000\pi\sigma^3/6$, $\kappa\equiv k\sigma^2/T_{\rm max}$ with time discretization $dt=2\times 10^{-3} \times\sigma^2/D_{\rm max}$, where the quantities $\mathcal{X}_{\rm max}=\max (\mathcal{X}_A,\mathcal{X}_B)$. This provides sufficiently large systems. We record the MSD(t) of each particle species in each realization after letting the system thermalize during a time $t \gtrsim 10^3-10^4 \times dt$ (then setting $t=0$)  and obtain statistics over 100 realizations. We run the simulations for sufficiently large times  until the diffusion constant $\bar{D}_{\alpha}(t)$ reaches a plateau (Fig.1(a)). We then fit a curve of the form $\bar{D}_{\alpha}(t)=D_{\alpha}+(D^{\rm s}_{\alpha}-D_{\alpha})\left(1-e^{-(t/\tau)^{\gamma}}\right)$ from which we can determine the asymptotic value $\bar{D}_{\alpha}(\infty)=D^{\rm s}_{\alpha}$. This optimization depends on $\gamma$ and $D^{\rm s}_{\alpha}$, but the resulting value of  $D^{\rm s}_{\alpha}$ is almost insensitive to $\gamma$. We choose $\gamma=1/2$ \cite{cichocki1990dynamic} for better numerical stability. Results  for a tagged particle in a bath (Fig.1) are obtained for $c_A=0.001$ in order to collect more data while for mixtures at finite concentration (Fig.2) we increased the volume size $V\rightarrow 2V$. 

We also determined indirect observables from the simulations. For a given parameter set $\Phi_{\rm }$ that includes concentrations, temperatures and sizes of the particles, the long-time friction constant is obtained from
\begin{equation}
\zeta^{\rm L}_{\alpha}(\Phi)/\zeta_{\alpha}=D_{\alpha}/D_{\alpha}^{\rm s}(\Phi_{\rm eq}),
\end{equation}
where $D_{\alpha}^{\rm s}(\Phi_{\rm eq})=D_{\alpha}^{\rm s}(c_{A},c_{B},\sigma_A,\sigma_B;T_{A}=T_{B})$ corresponds to the equilibrium value of self-diffusion constant. This way, we obtain effective temperatures from simulations for any mixture containing particles at temperatures $T_{A}$, $T_{B}$ by using
\begin{equation}
    T^{\text{eff}}_{\alpha}/T_{\alpha}=D_{\alpha}^{\rm s}(\Phi)/D_{\alpha}^{\rm s}(\Phi_{\rm eq})
\end{equation}
to compare with the theoretical result of Eq.(7) in the main text.

\appendix
\section{Derivation of averaging in Eq.\eqref{eq13}} \label{appendixa}
The fact that Eq.\eqref{eq13} requires right averaging is noted in Ref. \cite{leegwater1992dynamical}, and obviously is used in Ref. \cite{hanna1982self}. We have not found any explicit proof of this relationship, and hence we decided to add this part for pedagogical purposes and completeness.

The tagged-particle scattering function is the Fourier transform of the tagged-particle density autocorrelation function and is given by
\begin{equation}
F({\bf k},t)=\int e^{-i{\bf k}\cdot ({\bf r}(t)-{\bf r}(0))} p({\bf X}_t | {\bf X}_0) P^{\rm ss} ({\bf X}_0) d{\bf X}_0 d {\bf X}_t,
\end{equation}
where ${\bf r}_1(0)$ and ${\bf r}_1(t)$ are the positions at the initial time $t=0$ and at a later time $t$ of the tagged particle, ${\bf X_0}, {\bf X_t}$ are the sets of all particle coordinates at initial time and at time $t$, i.e., ${\bf X_0}\equiv \{{\bf r}_1(0),{\bf r}_2(0),...,{\bf r}_N (0) \}$, ${\bf X_t}\equiv \{{\bf r}_1(t),{\bf r}_2(t),...,{\bf r}_N (t) \}$, $p({\bf X}_t | {\bf X}_0)$  is the conditional probability of particles being at positions ${\bf X}_t$ at  time $t$   given that they were initially at ${\bf X}_0$, and $P^{\rm ss} ({\bf X}_0)$ is the steady-state probability distribution. 
We may write 
\begin{equation}
p({\bf X}_t | {\bf X}_0)=\int \delta ({\bf X}-{\bf X}_t)e^{\mathcal{L}_{\text{N}} t}\delta ({\bf X}-{\bf X}_0) d{\bf X},
\end{equation}
where $\mathcal{L}_{\text{N}}$ is the $N$-particle operator, ${\bf X}\equiv \{{\bf r}_1,{\bf r}_2,...,{\bf r}_N  \}$ and $\delta({\bf X}-{\bf X}_{t})=\prod_{i=1}^N \delta ({\bf r}_i-{\bf r}_i(t))$.
We can then perform the integral in $F({\bf k},t)$ over ${\bf X}_t$ to obtain:
\begin{equation}
F({\bf k},t)=\int P^{\rm ss} ({\bf X}_0) e^{-i{\bf k}\cdot ({\bf r}_1-{\bf r}_1(0))} e^{\mathcal{L}t}\delta ({\bf X}-{\bf X}_0) d{\bf X} d {\bf X}_0.
\end{equation}
Now, we may introduce the adjoint form in order to carry the integral over $d{\bf X}_0$, and hence we first rewrite:
\begin{equation}
F({\bf k},t)=\int P^{\rm ss} ({\bf X}_0) \delta ({\bf X}-{\bf X}_0) e^{i{\bf k}\cdot {\bf r}_1(0)} e^{\mathcal{L}_{\text{N}}^{\dagger} t} e^{-i{\bf k}\cdot {\bf r}_1} d{\bf X} d {\bf X}_0,
\end{equation}
where $\mathcal{L}^{\dagger}$ is the adjoint Liouvillian operator. Then, we perform the integral over ${\bf X}_0$ using $\delta({\bf X}-{\bf X}_{0})=\prod_{i=1}^N \delta ({\bf r}_i-{\bf r}_i(0))$ and obtain:
\begin{equation}
F({\bf k},t)=\int P^{\rm ss} ({\bf X})  e^{i{\bf k}\cdot {\bf r}_1} e^{\mathcal{L}_{\text{N}}^{\dagger} t} e^{-i{\bf k}\cdot {\bf r}_1} d{\bf X}.
\end{equation}
Finally, by reverting the adjoint form, we recover Eq.\eqref{eq13}, i.e.,
\begin{equation}
F({\bf k},t)=\int  e^{-i{\bf k}\cdot {\bf r}_1} e^{\mathcal{L}_{\text{N}} t} e^{i{\bf k}\cdot {\bf r}_1} P^{\rm ss} ({\bf X})  d{\bf X}=\langle e^{-i {\bf k}\cdot {\bf r}_1} e^{\mathcal{L}_{\rm N}t} e^{i {\bf k}\cdot {\bf r}_1}\rangle_{\text{ss}}.
\end{equation}

\end{document}